# Quantum State Transfer with Untunable Couplings

P. K. Gagnebin[1], S. R. Skinner[1], E. C. Behrman[2] and J. E. Steck[3]

We present a general scheme for implementing bi-directional quantum state transfer in a quantum swapping channel. Unlike many other schemes for quantum computation and communication, our method does not require qubit couplings to be switched on and off. The only control variable is the bias acting on individual qubits. We show how to derive the parameters of the system (fixed and variable) such that perfect state transfer can be achieved. Since these parameters vary linearly with the pulse width, our scheme allows flexibility in the time scales under which qubits evolve. Unlike quantum spin networks, our scheme allows the transmission of several quantum states at a time, requiring only a two qubit separation between quantum states. By pulsing the biases of several qubits at the same time, we show that only eight bias control lines are required to achieve state transfer along a channel of arbitrary length. Furthermore, when the information to be transferred is purely classical in nature, only three bias control lines are required, greatly simplifying the circuit complexity.

## I. INTRODUCTION

Currently, there are several ways of moving data around in a quantum computer. While some methods transfer quantum states by moving them down a linear array of qubits, there are others which exploit the quantum property of entanglement for teleporting quantum states between distant qubits [1, 2]. Recently, Bose proposed a scheme for using an unmodulated and unmeasured spin chain as a channel for short distance quantum communication [3]. The quantum state to be transmitted is placed on a spin at one end of the chain and is propagated to a spin at the other end by evolution under a suitable time-dependent Hamiltonian, without any additional external control. This makes these spin chains appealing because achieving dynamical control of the interactions between qubits is often a problem encountered in quantum computer implementations [4, 5]. Moreover, for short-length chains, the fidelity of state transfer is close to unity. However, due to dispersion of the initial information over the entire length of the chain, as the length of the chain increases, the fidelity of state transfer substantially reduces. To overcome this problem, several schemes to achieve perfect state transfer have been proposed. One of these schemes uses spatially varying coupling constants to refocus the information at the receiving end of the chain [6-8]. Another method is to encode the information in low-dispersion Gaussian wave packets spread over several spins [9]. Very high fidelity can also be achieved in chains where the first and last qubits are only weakly coupled to the rest of the chain [10]. Conclusive state transfer is another promising method where parallel quantum channels are used, which in addition to providing high fidelity state transfer, are more robust to decoherence and non-optimal timing than the single chain schemes [11, 12].

The major advantage of using quantum spin chains for state transfer is their simplicity. However, since the state initialized to the first qubit of the chain propagates to the last qubit after a *certain* time duration governed by its evolution under the Hamiltonian, only a single state can be transferred at a time. This places a limit on the rate at which quantum information is transferred. In a quantum computer involving several qubits, transporting several states down a channel might be more efficient. In this paper, we present a scheme for transferring quantum information along a linear arrangement of qubits with nearest-neighbor interactions, whereby several quantum states can be transported at a time from one end of the wire to another. Since most proposals for solid-state implementations of a quantum computer use a one-dimensional line of qubits with nearest-neighbor interactions [13-22], our scheme can be used to implement a quantum wire in these applications.

As with the spin channels, we achieve state transfer without having to switch "on" and "off" the coupling between adjacent qubits. Recently, Zhou *et al.* devised a scheme for universal and scalable quantum computation without the need to tune the couplings between qubits [4]. Their method relies on the idea of computing with logical bits, which comprise several physical bits [23]. The coupling between the encoded qubits are effectively turned on and off by computing in and out of carefully designed interaction free subspaces analogous to decoherence free subspaces [24, 25]. Our scheme, unlike theirs, transfers the states of physical qubits and not of encoded qubits where state transfer is achieved by means of swap operations between adjacent qubits. In a one-dimensional arrangement of qubits with nearest neighbor interactions, each qubit is coupled to a qubit on either side of it. Therefore, when a gate operation is performed on a qubit, the evolution of the three-qubit system is governed by an $8 \times 8$ Hamiltonian matrix. We show how this Hamiltonian can be reduced to a $2 \times 2$ matrix describing the evolution of the "target" qubit only, by fixing the states of the two qubits coupled to it. The evolution of the target qubit can then be described as taking place in different two-dimensional subspaces of the eight-dimensional Hilbert space of the three qubits and this reduction technique can be used to solve for parameters (bias, tunneling and coupling) in order to realize swap operations between two adjacent qubits. The tunneling and coupling parameters obtained in our solutions are "fixed" parameters of the system; the only control parameter is the bias acting on individual qubits. We further show that by pulsing the bias on alternate qubits at the same time, only eight bias control lines are required for a quantum channel of arbitrary length.

Another advantage of our scheme is that the time scales under which qubits evolve are flexible and can, therefore, be adjusted to the requirements of the particular experimental realization. This is because the governing equations used to solve for the parameters are scalable and depending on the time duration of the applied bias pulse, the parameters can be scaled. One of the drawbacks of spin channels using constant coupling is that the time required for transfer is long compared to qubit decoherence times, making state readout and manipulation impossible using current experimental technology [26, 27]. By dynamically varying the coupling this problem can be solved as shown by Lyakhov and Bruder [28], in which they vary the coupling between the first and the last pair of qubits. In practical quantum computing applications, varying the coupling between qubits might not always be possible. For instance, in Josephson junction devices [29-36], the coupling is usually realized using a hard-wired capacitor or inductor which is fixed during fabrication and therefore, cannot be tuned during computation. Even though a number of variable coupling schemes [32, 34-36] have been devised, they are not completely satisfactory. Most of them require external controls making them major decoherence sources [32, 34], while others avoiding the use of external controls in their design are limited in the number of qubits that can be incorporated [35, 36]. Therefore, a scheme which allows state transfer without switching the couplings is very useful because, besides reducing decoherence, it simplifies the operation drastically.

## II. QUANTUM SWAPPING CHANNEL

Consider a linear arrangement of qubits, Figure 1, where each qubit is represented as a circle. We assume weak coupling between the qubits and each qubit is coupled only to the qubits adjacent to it through the coupling terms, $\xi$, represented in the figure by square boxes. The coupling between adjacent qubits is a fixed parameter of the system. The only variable is the bias operating on individual qubits which will be pulsed between high and low values in order to achieve a quantum wire operation. We assume all the qubits are identical in having the same value for the tunneling parameter. In subsequent sections, we will show how to calculate the parameter values of the system, both fixed (coupling and tunneling) and variable (bias), such that a quantum wire operation is realized.

Quantum state transfer is achieved by swapping the states of adjacent qubits and thereby moving states along the line of qubits from one end to the other. Conventionally, a swap operation between two qubits comprises three Controlled-NOT (CNOT) gate operations. Previously [37], we have showed that a CNOT gate operation between two qubits can be realized in a single pulse operation by pulsing the bias on the target qubit only to a certain "low" value for a certain time duration. Figure 1 shows the sequence of CNOT pulses applied on two qubits, B and C, involved in a swap operation. The first CNOT pulse at time $t_1$ is applied to qubit B. Next, a CNOT pulse is applied to qubit C at time $t_2$ and finally a CNOT pulse is applied to qubit B again at time $t_3$. At the end of the third CNOT pulse, the states of qubits B and C are interchanged.

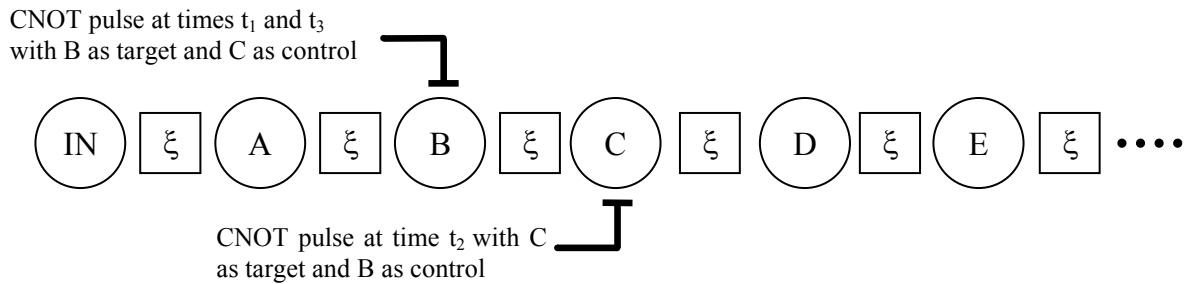

Fig. 1. Quantum swapping channel comprising a linear arrangement of qubits. Each qubit is coupled to the qubits adjacent to it through the coupling terms, $\xi$. Here, a swap operation is performed between qubits B and C where three CNOT pulses are applied successively to the qubits. The first CNOT pulse is applied to qubit B at time $t_1$, the next CNOT pulse is applied to qubit C at time $t_2$ and the last CNOT pulse is applied to qubit B at time $t_3$.

Since in our scheme the coupling between adjacent qubits is permanently "on", qubits A and D are always coupled to qubits B and C, respectively. Since qubits A and D can be in any *arbitrary* quantum state, the evolution of the two-qubit system comprising qubits B and C depends on the states of qubits A and D and will vary as the states of these two qubits vary. The question arises whether the system comprising qubits B and C can be isolated from qubits A and D without having to switch off the coupling between qubits A and B and between qubits C and D. One approach is to initialize qubits A and D to the $|0\rangle$ state. This will allow us to *exactly* account for the effect of the coupling term between qubits A and B (C and D) as we will subsequently show.

Suppose the system of three qubits A, B and C is described by the following $8 \times 8$ Hamiltonian:

$$H = \Delta\sigma_{XA} + \Delta\sigma_{XB} + \Delta\sigma_{XC} + \varepsilon_A\sigma_{ZA} + \varepsilon_B\sigma_{ZB} + \varepsilon_C\sigma_{ZC}$$
$$+ \xi_{AB}\sigma_{ZA}\sigma_{ZB} + \xi_{BC}\sigma_{ZB}\sigma_{ZC} \quad (1)$$

where $\Delta$ is the tunneling parameter of a qubit, $\varepsilon_A$, $\varepsilon_B$ and $\varepsilon_C$ are the biases acting on qubits, A, B and C, respectively, $\xi_{AB}$ and $\xi_{BC}$ are the coupling constants between qubits A and B and between qubits B and C, respectively, which are each equal to "$\xi$" (Fig. 1), $\sigma_{ZA}$, $\sigma_{ZB}$ and $\sigma_{ZC}$ are the Pauli spin matrices for qubits A, B and C, respectively. Note that we have not considered qubit D in the system. This is because under a CNOT gate operation we will be varying the bias only on the target qubit (B in this case) which is coupled only to two qubits (A and C). As shown in [37], by maintaining high biases on qubits A and C whereby their states are "fixed", the 8 × 8 matrix as given by Eq. (1) can be reduced to a 2 × 2 matrix describing the evolution of the target qubit B:

$$H_B = \Delta_B\sigma_X + (\varepsilon_B \pm \xi_{BC} + \xi_{AB})\sigma_Z \quad (2)$$

The coupling term $\xi_{BC}$ either adds or subtracts from the bias term $\varepsilon_B$ depending on whether the control qubit C is in the $|0\rangle$ or $|1\rangle$ state, respectively. This is because the expectation value of the $\sigma_{ZC}$ operator in the subspace of qubit B is +1 or -1 depending on whether qubit C is in the $|0\rangle$ or $|1\rangle$ states, respectively. Since we initialized qubit A to the $|0\rangle$ state, the expectation value of the $\sigma_{ZA}$ operator in the subspace of qubit B is always +1. Therefore, the coupling term $\xi_{AB}$ always adds to the bias term $\varepsilon_B$.

Starting in an initial state, say $|0\rangle$, the probability of qubit B in the $|1\rangle$ state can be written as an oscillatory function in terms of the parameters $\Delta$ (tunneling) and $\varepsilon_B$ (bias) as follows [37]:

$$P_{|1\rangle} = X - Y\cos(2\pi ft), \quad (3)$$

where the offset $X$, the amplitude $Y$ and the frequency $f$ of probability oscillation are given as:

$$\text{Offset, } X = \left(\frac{1}{2} - \frac{(\varepsilon \pm \xi + \xi)^2}{2(\Delta^2 + (\varepsilon \pm \xi + \xi)^2)}\right), \quad (4)$$

$$\text{Amplitude, } Y = \frac{\Delta^2}{2(\Delta^2 + (\varepsilon \pm \xi + \xi)^2)}, \quad (5)$$

$$\text{Frequency, } f = 2\sqrt{\Delta^2 + (\varepsilon \pm \xi + \xi)^2}, \quad (6)$$

Here, we have replaced the bias term $\varepsilon_B$ with $\varepsilon$ and the terms $\xi_{AB}$ and $\xi_{BC}$ by $\xi$. We have chosen a basis where the Planck's constant is normalized to 1. From Eq. (6), there will be two frequencies of oscillation for the probability function. By making the bias term equal to zero, these two frequencies of oscillation will be given by:

$$f_1 = 2\sqrt{\Delta^2 + 4\xi^2} \quad (7)$$

$$f_2 = 2\sqrt{\Delta^2} = 2\Delta \quad (8)$$

The frequencies $f_1$ and $f_2$ correspond to the cases when the control qubit C is in the $|0\rangle$ and $|1\rangle$ states, respectively. Under a CNOT gate operation, we require that the target qubit flip its state when the control qubit is in the $|1\rangle$ state and that the target qubit remain in its state when the control qubit is in the $|0\rangle$ state. Given a time step, $T$, of pulse operation, *we therefore require frequency $f_1$ to correspond to an integer multiple of complete oscillations and frequency $f_2$ to correspond to an odd integer multiple of half-cycle oscillations* [37]. Under these conditions, for a chosen time step $T$, using the conditions imposed on frequencies $f_1$ and $f_2$ to realize a CNOT gate, we have a system of two equations in two unknown parameters, $\Delta$ and $\xi$, as follows:

$$f_1 = 2\sqrt{\Delta^2 + 4\xi^2} = \frac{M}{T} \quad (9)$$

$$f_2 = 2\Delta = \frac{(2N+1)/2}{T} \quad (10)$$

where $M$ and $N$ are integers. The parameters, $\Delta$ and $\xi$, can therefore be solved for and these form the fixed parameters of the system. For instance, for a time step of 10ns, we obtain values of 25 GHz and 21.65 MHz for the tunneling and coupling, respectively, which are experimentally realizable values for rf-SQUIDs [37]. Recall that we had previously chosen the value of the bias, $\varepsilon$, to be zero. Therefore, when we apply a pulse to the target qubit, the bias on it will be pulsed from a high value to zero and will be maintained at this value for the entire time step, $T$. We call this zero-bias pulse a CNOT pulse. The high value of the pulse can be arbitrarily chosen as long as it is sufficiently higher than the tunneling parameter, $\Delta$. It is important to note that these parameters which we solve for using Eqs. (9) and (10) will realize a CNOT gate operation between the two adjacent qubits B and C in a three-qubit coupled system of qubits A, B and C, provided the state of qubit A is "fixed" to the $|0\rangle$ state. Also, instead of choosing a time step $T$, we can choose a value for the tunneling parameter (or the coupling) depending upon the physical system under consideration and solve for the time step $T$ and the coupling (or the tunneling).

Therefore, by fixing the state of qubit A, we are able to realize a CNOT gate operation in a three qubit system analogous to realizing such a gate operation in a two qubit system. We introduce the term "sacrificial" qubit for qubits like A whose state is known (fixed to the $|0\rangle$ state) and therefore, whose presence can be accounted for in solving for the system parameters.

Next, when a CNOT pulse ("zero" bias pulse) is applied to qubit C, with qubit B as the control, qubit D will be treated as a sacrificial qubit by initializing its state to the $|0\rangle$ state. The key point to observe is that when a swap operation is performed between any two qubits, the qubits adjacent to them must be initialized to the $|0\rangle$ state and therefore, treated as sacrificial qubits. In the next section, we will describe an efficient scheme for implementing the quantum swapping channel making use of sacrificial qubits in our design.

The parameters derived for realizing a CNOT gate operation using our scheme in fact realizes the CNOT gate modulo a phase shift. When the control qubit is in the $|0\rangle$ and $|1\rangle$ states, phase shifts of $\pi$ and $-\pi/2$, respectively, are introduced. The following transformations occur under the CNOT gate operation:

$$|00\rangle \to \exp(i\pi)|00\rangle \,;\, |01\rangle \to \exp(i\pi)|01\rangle \,;$$

$$|10\rangle \to \exp(-i\pi/2)|11\rangle \,;\, |11\rangle \to \exp(-i\pi/2)|10\rangle. \quad (11)$$

Therefore, each swap operation introduces additional phases on the quantum states being swapped. If at the end of the swapping channel, the quantum data is being measured this additional phase would not be of any consequence. However, if the quantum states transported down the swapping channel are to be used for quantum computing making use of quantum properties like interference, the additional phases would affect the computation. It will, thus, become important to devise a scheme for either eliminating these phases or accounting for them. We will show how the scheme implemented by us in the next section totally eliminates any phases by the time a quantum state reaches the end of the swapping channel.

## III. IMPLEMENTING THE QUANTUM SWAPPING CHANNEL

In the previous section, we stated that when two neighboring qubits are involved in a swap operation, the qubits adjacent to them must be treated as sacrificial qubits with their states initialized to the $|0\rangle$ state. Therefore, the role of sacrificial qubits is not fixed to certain qubits in the design. Depending on where swap operations are performed, different qubits will be assigned the role of sacrificial qubits.

Figure 2 shows a design scheme for implementing a quantum swapping channel where only five qubits have been considered to demonstrate the operation. Following usual conventions in the literature, each of the five horizontal lines represents a wire carrying a single qubit and time goes from left to right. The qubits have been labeled on the left hand side of each horizontal line. The states of the qubit before and after an operation is performed on it are represented above the line corresponding to the qubit. The two x-marks connected by a vertical line represent a swap operation (3 CNOT pulses) and the rectangular box with $|D_i\rangle$ written in it represents an initialization of the qubit to the arbitrary quantum state $|D_i\rangle$. The rectangular box with R written in it represents reading out the state of the qubit and re-initializing it to the $|0\rangle$ state.

Suppose that initially the states of the qubits IN through OUT are $|D_1\rangle, |0\rangle, |0\rangle, |D_0\rangle, |0\rangle$, respectively, where $|D_0\rangle$ and $|D_1\rangle$ are arbitrary quantum states and represent data being transmitted along the wire. (In a practical implementation, all the qubits except IN will be initialized to the $|0\rangle$ state and qubit IN will be prepared in an arbitrary quantum state $|D_0\rangle$. By performing successive swap operations in the sequence shown in Fig. 2, the system eventually will evolve to the state $|D_1\rangle, |0\rangle, |0\rangle, |D_0\rangle, |0\rangle$. We have not showed these initial operations here due to space considerations). Swap operations are performed between qubits IN and A and between qubits C and OUT. Notice that qubit B acts as a sacrificial qubit since it is in the $|0\rangle$ state. Each swap operations comprise three CNOT pulses (Fig. 1) where the bias on the target qubit is pulsed to zero for a time step $T$. In this case, the first pulse is applied to qubits IN and C, the second pulse is applied to qubits A and OUT and the third pulse is applied to qubits IN and C again. The states of qubits IN through OUT after the swap operation are $|0\rangle, |D_1\rangle, |0\rangle, |0\rangle$ and $|D_0\rangle$, respectively. Notice that the state of qubit OUT is available for read-out, and after being read, it is re-initialized to the $|0\rangle$ state. Simultaneously, swap operations are performed between qubits A and B with qubits IN and C acting as sacrificial qubits. The states of the qubits IN through OUT after these operations are $|0\rangle, |0\rangle, |D_1\rangle, |0\rangle$ and $|0\rangle$, respectively. Next, a swap operation is performed between qubits B and C with qubits A and OUT acting as sacrificial qubits. Simultaneously, qubit IN is prepared in the next input state, $|D_2\rangle$. The states of the qubits IN through OUT now are $|D_2\rangle, |0\rangle, |0\rangle, |D_1\rangle$ and $|0\rangle$, respectively. This process of performing swap operations on qubits, each time treating the qubits adjacent to them not involved in the swap as sacrificial qubits, is carried on in order to realize a quantum swapping channel. Notice that, unlike the others, qubits IN and OUT are each coupled to a single qubit. Therefore, under a CNOT gate operation when either of these two qubits function as targets, the bias on them is not pulsed to zero, but to a value equal to the coupling term, $\xi$, as described in Reference [37]. In other words, a usual two-qubit CNOT gate is realized when either of these two qubits act as target qubits.

In our scheme, we are assuming that the time taken to prepare qubit IN in a given quantum state $|D_i\rangle$, the time taken to read out the state of qubit OUT and to initialize it to the $|0\rangle$ state is comparable to the time taken to do a swap operation. This assumption can be supported by the fact that the parameter values for which a CNOT gate is realized vary linearly with the time step. For instance, if the time step were changed from 10ns to 100ns, the values of $\Delta$ and $\xi$ would change from 25 MHZ and 21.65 MHZ to 2.5 MHZ and 2.165 MHZ, respectively. Therefore, for a particular experimental realization the parameters can be appropriately scaled depending on the time step under consideration.

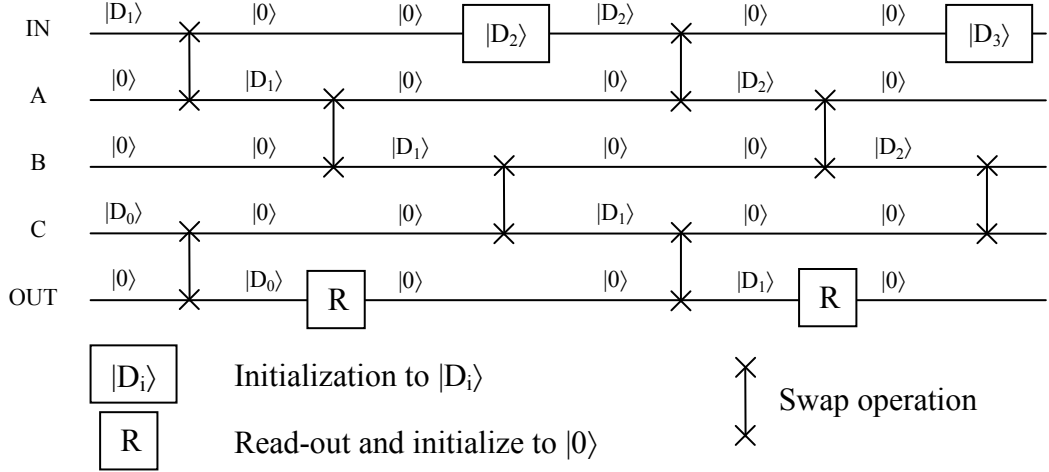

Fig. 2. Pulsed bias sequences on a linear arrangement of five qubits IN, A, B, C and OUT, comprising a quantum wire. Each of the five horizontal lines represents a wire carrying a single qubit and time goes from left to right. The qubits have been labeled on the left hand side of each horizontal line. The states of the qubit before and after an operation is performed on it are represented above the line corresponding to the qubit. The two x-marks connected by a vertical line represent a swap operation (3 CNOT pulses) and the rectangular box with $|D_i\rangle$ written in it represents an initialization of the qubit to the state $|D_i\rangle$, which is an arbitrary quantum state. The box with R represents reading out the state of the qubit and re-initializing it to the $|0\rangle$ state. Note that in our design implementation scheme, only eight bias control lines are required, taking into account that each swap operation requires two control lines and qubits IN and OUT require separate bias control lines since the bias on these qubits is pulsed to the value of the coupling term and not to zero.

At the end of the previous section, we mentioned how additional phases are introduced in the quantum states being swapped as a result of each gate operation realizing the CNOT gate modulo a phase shift (Eq. (11)). We show here how our scheme allows us to cancel out all phases by the time a particular quantum state is transported to the end of the channel. Suppose the initial state $|D_1\rangle$ in Fig. 2 is an arbitrary quantum state $\alpha|0\rangle + \beta|1\rangle$ where $|\alpha|^2 + |\beta|^2 = 1$. Under the first swap operation, 3 CNOT pulses are applied to qubits IN, A and IN, respectively. The equations below shows the transformation of the two qubits along with the phases introduced as given by Eq. (11) where the first and second qubits represent qubits IN and A, respectively.

$$\alpha|00\rangle + \beta|10\rangle \rightarrow \exp(i\pi)\alpha|00\rangle + \exp(-i\pi/2)\beta|11\rangle$$
$$\rightarrow \exp(i\pi + i\pi)\alpha|00\rangle + \exp(-i\pi/2 - i\pi/2)\beta|01\rangle$$
$$\rightarrow \exp(i\pi + i\pi + i\pi)\alpha|00\rangle + \exp(-i\pi/2 - i\pi/2 + i\pi)\beta|01\rangle$$
$$= \exp(i\pi)\alpha|00\rangle + \beta|01\rangle = |0\rangle(\exp(i\pi)\alpha|0\rangle + \beta|1\rangle)$$
(13)

From Eq. (13) it is clear that only the $|0\rangle$ state of the quantum state $\alpha|0\rangle + \beta|1\rangle$ picks up an additional phase of $\pi$. Therefore, if even numbers of swap operations are performed in transferring the state from qubits IN to OUT, the quantum state $\alpha|0\rangle + \beta|1\rangle$ can be exactly transported down the swapping channel without introducing an additional relative phase between the $|0\rangle$ and $|1\rangle$ states. This is possible if the swapping channel is built using an *odd* number of qubits. In Fig. 2, we use 5 qubits whereby 4 swap operations are performed on a quantum state in transporting it from qubits IN to OUT.

It is worth pointing out that in our design implementation scheme, only eight bias control lines are required, taking into account that each swap operation requires two control lines and qubits IN and OUT require separate bias control lines since the bias on these qubits is pulsed to the value of the coupling term and not to zero. We always require only eight bias control lines, no matter how many qubits are involved in the design of the quantum wire, assuming at least 5 qubits are used. This is, therefore, an efficient implementation since the number of control lines is minimal. Moreover, due to symmetry of the architecture, the quantum swapping channel is bi-directional.

When the data states to be transferred down the line of qubits are purely classical in nature, i.e., the qubits are always only in one of the two basis states, $|0\rangle$ or $|1\rangle$, the number of gate operations can be further reduced. In this case, we can move the qubit states from one end of the channel to the other by copying the state of the preceding qubits onto the qubit adjacent to it. This is because the no-cloning theorem [38] only forbids the copying of arbitrary quantum states and not the copying of a qubit in a known state. In the next section, we show how the number of gate operations can be reduced in transmitting only basis states down the line of qubits. Moreover, we show that only three bias control lines are required in this case.

## IV. MOVING CLASSICAL DATA

Figure 3 shows a schematic of the quantum wire shown in Fig. 1 with the qubits renamed as IN, $A_1$, $B_1$, …, $A_N$, $B_N$, OUT (There are $2N+2$ qubits in all, where N is an integer). As before, the quantum wire operation is performed by applying a sequence of bias pulses. However, in this case, only three bias lines $\phi_1$, $\phi_2$ and $\phi_3$ are required (Fig. 3) instead of eight. Bias line $\phi_1$ is connected to the bias lines of the N qubits $A_1$, $A_2$, …, $A_N$, while bias line $\phi_2$ is connected to the bias lines of the N qubits $B_1$, $B_2$, …, $B_N$. The clock pulses $\phi_1$ and $\phi_2$ only differ in phase and the qubits they are applied to. Bias line $\phi_3$ is used to pulse the bias on the output qubit OUT to move data out of the wire and is simultaneously applied with $\phi_1$.

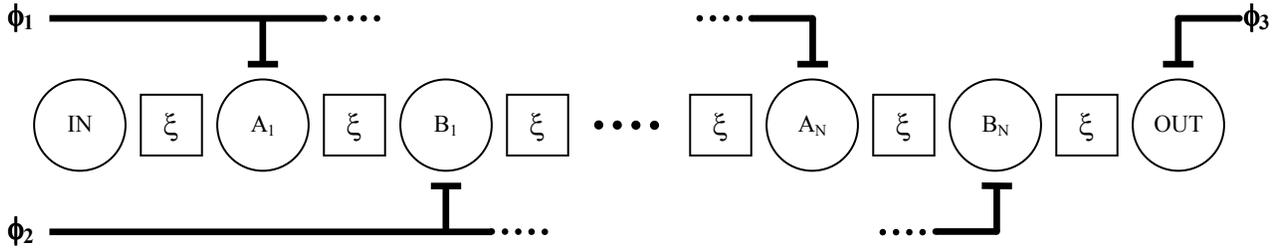

Fig. 3. Quantum wire using a linear arrangement of qubits IN through OUT similar to Fig. 1. Data is entered for a shift operation by preparing qubit IN in the desired input state and then copying its state onto successive qubits. There are three bias lines - $\phi_1$ used to pulse the bias on alternate qubits $A_1$ through $A_N$ simultaneously, $\phi_2$ used to pulse the bias on alternate qubits $B_1$ through $B_N$ simultaneously, and $\phi_3$ used to pulse the bias on qubit OUT, which is applied simultaneously with $\phi_1$.

Classical data is moved down the line of qubits by copying the state of the preceding qubit onto the qubit to which a pulse is applied. We call this pulse a COPY pulse which constitutes the pulse applied using bias lines $\phi_1$ and $\phi_2$. Table I shows a general state table of the system of three qubits (in this case, IN, $A_1$ and $B_1$) before and after a COPY pulse is applied to the target qubit $A_1$. Note that the initial states of qubits $A_1$ and $B_1$ are the same. This is because prior to the application of a COPY pulse to qubit $A_1$, a COPY operation was performed on qubit $B_1$ where the state of qubit $A_1$ was copied to it, assuming the direction of shift of data states is from left to right. From the table we can see that under a COPY pulse, the target qubit flips its state only when both the control qubits adjacent to it are in opposite states. When the two control qubits are in the same state, the target qubit does not change its state.

TABLE I. State table showing the states of qubits IN, $A_1$ and $B_1$, before and after a COPY pulse is applied to qubit $A_1$. Both qubits $A_1$ and $B_1$ are initially in the same state ($|0\rangle$ or $|1\rangle$). On applying a COPY pulse to qubit $A_1$, it flips its state only when qubits IN and $B_1$ are in opposite states.

| Initial states | | | Final states | | |
|---|---|---|---|---|---|
| IN | $A_1$ | $B_1$ | IN | $A_1$ | $B_1$ |
| $|0\rangle$ | $|0\rangle$ | $|0\rangle$ | $|0\rangle$ | $|0\rangle$ | $|0\rangle$ |
| $|0\rangle$ | $|1\rangle$ | $|1\rangle$ | $|0\rangle$ | $|0\rangle$ | $|1\rangle$ |
| $|1\rangle$ | $|0\rangle$ | $|0\rangle$ | $|1\rangle$ | $|1\rangle$ | $|0\rangle$ |
| $|1\rangle$ | $|1\rangle$ | $|1\rangle$ | $|1\rangle$ | $|1\rangle$ | $|1\rangle$ |

As described in Section II, by maintaining a high bias on the two qubits IN and $B_1$, the $8 \times 8$ Hamiltonian describing the evolution of qubits IN, $A_1$ and $B_1$ which is of the form of Eq. (1) can be reduced to a $2 \times 2$ matrix describing the evolution of qubit $A_1$ only. The reduced $2 \times 2$ Hamiltonian matrix for the target qubit $A_1$ interacting with the two control qubits, IN and $B_1$, can therefore be written as

$$H_{A1} = \Delta \sigma_X + \Sigma \sigma_Z \text{ with } \Sigma = \varepsilon \pm \xi \pm \xi \quad (12)$$

where $\Sigma$ is the effective bias acting on the target qubit as a result of the coupling terms either adding or subtracting from the bias applied to the target qubit $A_1$, depending on the states of the two adjacent qubits, IN and $B_1$. Note here that both qubits IN and $B_1$ function as controls and neither acts as a sacrificial qubit. Analogous to Eq. (3), the probability of the target qubit in the $|1\rangle$ state can be written as an oscillatory function of time with three different frequencies of oscillation (as a result of the three different effective biases). The three frequencies of oscillation are:

$$f_1 = 2\sqrt{\left(\Delta^2 + (\varepsilon + 2\xi)^2\right)}, \; f_2 = 2\sqrt{\left(\Delta^2 + \varepsilon^2\right)}, \text{ and}$$

$$f_3 = 2\sqrt{\left(\Delta^2 + (\varepsilon - 2\xi)^2\right)}, \quad (13)$$

where $f_1$ and $f_3$ correspond to the cases when both qubits IN and $B_1$ are in the $|0\rangle$ and $|1\rangle$ states, respectively, and $f_2$ corresponds to the case when they are in opposite states. By choosing the bias term equal to zero during the COPY pulse, we can have two frequencies of oscillation as follows:

$$f_1 = 2\sqrt{(\Delta^2 + 4\xi^2)}, \qquad (14)$$

$$f_2 = 2\Delta. \qquad (15)$$

These are the same frequencies given by equations (7) and (8) even though the interpretations of these frequencies under the two cases are different. In this case, frequencies $f_1$ and $f_2$ correspond to the frequencies of oscillation when the control qubits are in the same state and in opposite states, respectively. Therefore, if $T$ is the time duration for which the COPY pulse is applied to qubit $A_1$, *frequency $f_2$ must be chosen such that an odd integer number of half cycles is realized within the time step, T*. This will cause the target qubit to flip its state when the two control qubits are in opposite states, in accordance with Table I. When the control qubits are in the same state, we require qubit $A_1$ to maintain its state. Therefore, *frequency $f_1$ must be chosen to correspond to an integer number of complete oscillation cycles, within the time step T of the pulse operation*. Notice that we have arrived at the same conditions to be satisfied for frequencies $f_1$ and $f_2$ as that required for implementing a quantum swapping channel in Section II. Therefore, the same parameters used to realize a CNOT pulse can be used to realize a COPY pulse within the same time step. Similar to a CNOT pulse when a COPY pulse operation is performed on qubit $A_1$, its bias is pulsed to "zero" for a time step $T$.

To copy the state of qubit $B_N$ onto qubit OUT, qubit OUT is first initialized to the $|0\rangle$ state and then a CNOT gate operation is performed between qubits $B_N$ and OUT with qubit OUT as the target. As mentioned in Section II, the bias on qubit OUT is pulsed to a value equal to the coupling parameter, $\xi$, between the two qubits for the same time step $T$ [37]. This bias pulse is applied to qubit OUT using bias line $\phi_3$ and we call it the READ-OUT pulse. The state of qubit OUT is now available for read-out and once it is read, qubit OUT is re-initialized to the $|0\rangle$ state before the next CNOT gate operation can be performed.

To demonstrate the classical wire operation, consider as an example a system of six qubits IN, $A_1$, $B_1$, $A_2$, $B_2$ and OUT, in an arrangement similar to that shown in Fig. 3. Figure 4 shows the pulse sequences on each of the five qubits, $A_1$ through OUT. Black solid dots and crosses are used to represent qubits acting as controls and targets, respectively, during a pulse operation. Each target qubit is only affected by the control qubit/s adjacent to it since we are only assuming nearest-neighbor interactions in our design. Note that the states of all the qubits, targets as well as controls, are depicted after a pulse is applied.

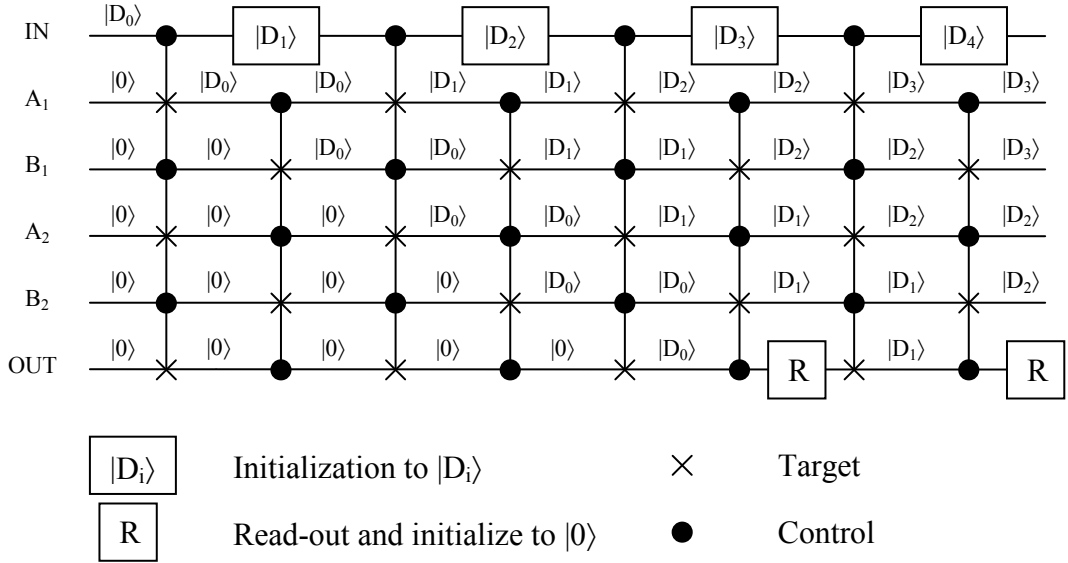

Fig. 4. Pulsed bias sequences to achieve a classical wire operation in an architecture comprising only six qubits, IN, $A_1$, $B_1$, $A_2$, $B_2$ and OUT. Each qubit is represented as a line and time goes from left to right. Alternate qubits are pulsed simultaneously. This reduces the number of control lines required to implement the operation. Vertical lines with dots and crosses represent pulses. During a pulse operation, control qubits are represented as dots and target qubits by crosses. Each target qubit is only affected by the control qubit/s adjacent to it. The state of each qubit after a pulse is applied is represented as $|D_i\rangle$, on the line corresponding to the qubit.

At the start of the shift operation, all the qubits $A_1$ through OUT are initialized to the $|0\rangle$ state. Qubit IN is initialized to the data state $|D_0\rangle$ which can be either of the two basis states. Now, the biases on qubits $A_1$, $A_2$ and OUT are simultaneously pulsed by applying a COPY pulse to bias line $\phi_1$ and a READ-OUT pulse to bias line $\phi_3$. From Fig. 3, we can see that after these pulses ($\phi_1$ and $\phi_3$) are applied, the states of qubits IN, $B_1$ and $B_2$ are copied onto qubits $A_1$, $A_2$ and OUT, respectively. Next, a COPY pulse is applied to bias line $\phi_2$, at the end of which the states of qubits $A_1$ and $A_2$ are copied onto qubits $B_1$ and $B_2$, respectively. Also, while pulse $\phi_2$ is applied, qubit IN is simultaneously prepared in the new state $|D_1\rangle$ as depicted by a box with the state "$|D_1\rangle$" in it. The process of applying pulses $\phi_1$ and $\phi_3$ simultaneously and then applying pulse $\phi_2$ is carried out sequentially to realize a classical wire operation. Note that after the third sequence of pulses is applied (a pulse sequence comprises a simultaneous application of pulses $\phi_1$ and $\phi_3$ followed by an application of pulse $\phi_2$), qubit OUT is in the state $|D_0\rangle$, the first data state transferred out of the channel. This is now available for read-out after which qubit OUT is re-initialized in the $|0\rangle$ state. In general, if the channel comprises 2N qubits, the first state qubit IN is prepared in is available for read-out after N sequences of clock pulses.

While the quantum swapping channel described in Section III can be used to move classical data, the scheme described in this section presents a better option. This is because unlike a swap operation which requires three CNOT pulses, a copy operation can be achieved in a single pulse. Also, only three bias control lines are required in this scheme as compared to the swapping channel where a minimum of eight lines are required. Therefore in moving classical data in a quantum computer, the scheme described in this section is more efficient. Transporting classical data using qubits will become essential when dealing with architectural issues of a heterogeneous quantum/classical computer as that discussed by Jonker and Han in [39].

In our scheme we have assumed that qubits tied to a common bias line are all identical, having the same parameter values. In other words, we have not taken into account the effect of errors due to slight mismatch of parameters during fabrication. Moreover, we have assumed ideal pulses in our derivations. It will be interesting to study the effect of mismatches in parameters, of finite rise and fall times, and of decoherence due to classical control on state transfer using this scheme. The design challenges to be addressed due to propagation of errors as a result of these mismatches will be pursued as a future work.

## V. CONCLUSION

We have shown in this paper a general approach for achieving state transfer in a quantum swapping channel without having to switch the coupling "off" between adjacent qubits. State transfer is achieved by varying the biases on individual qubits. However, since the biases of several qubits are pulsed at the same time, we require only eight bias control lines for a channel of arbitrary length. Therefore, the scheme is efficient. Furthermore, we also show that the number of bias control lines can be reduced from eight to three when the data to be transported is classical in nature. The main advantage of our scheme is that the time scales under which qubits evolve are flexible and, therefore, can be adjusted to the requirements of the particular experimental realization. This is because the governing equations used to solve for the parameters are scalable and depending on the time duration of the applied bias pulse, the parameters can be scaled. Moreover, transfer of quantum information is not restricted to a single state in our scheme and several qubits can be transported at the same time. This is because we do not wait for the state to evolve under the system Hamiltonian from the first qubit to the last along the chain as is the case in quantum spin chains.

## ACKNOWLEDGEMENT

This work was supported by the National Science Foundation, Grant # ECS- 0201995.

___________________________________________

[1]  N. Isailovic, M. Whitney, Y. Patel, J. Kubiatowicz, D. Copsey, F. T. Chong, I. L. Chuang and M. Oskin, ACM Trans. On Architecture and Code Optimization **1**, No. 1, 34 (2004).
[2]  M. Oskin *et. al*, "Building Quantum Wires : the Long and Short of it," In Computer Architecture News, Proc. 30th Annual International Symposium on Computer Architecture. ACM, June 2003.
[3]  S. Bose, Phys. Rev. Lett. **91**, 207901 (2003).
[4]  X. Zhou, Z. Zhou, G. Guo and M. J. Feldman, Phys. Rev. Lett. **89**, 197903 (2002).
[5]  S. C. Benjamin and S. Bose, Phys. Rev. Lett. **90**, 247901 (2003).
[6]  M. Christandl, N. Datta, A. Ekert and A. J. Landahl, Phys. Rev. Lett. **92**, 187902 (2004).
[7]  M. Paternostro, G. M. Palma, M. S. Kim and G. Falci, Phys. Rev. A **71**, 042311 (2004).
[8]  P. Karbach and J. Stolze, Phys. Rev. A **72**, 030301(R) (2005).
[9]  T. J. Osborne and N. Linden, Phys. Rev. A **69**, 052315 (2004).
[10] A. Wojcik, T. Luczak, P. Kurzynski, A. Grudka, T. Gdala and M. Bednarska, Phys. Rev. A **72**, 034303 (2005).
[11] D. Burgarth and S. Bose, Phys. Rev. A **71**, 052315 (2005).
[12] D. Burgarth, S. Bose and V. Giovannetti, J. Phys. A: Math. Gen. **38**, 6793 (2004).
[13] B. E. Kane, Nature (London) **393**, 133 (1998).
[14] A. J. Skinner, M. E. Davenport and B. E. Kane, Phys. Rev. Lett., **90**, 087901 (2003).

[15] R. Vrijen, E. Yablonovitch, K. Wang, H. W. Jiang, A. Balandin, V. Roychowdhury, T. Mor and D. DiVincenzo, Phys. Rev. A **62**, 012306 (2000).
[16] K. Yang et. al, Chinese Phys. Lett **20**, 991 (2003).
[17] J. K. Pachos and P. L. Knight, Phys. Rev. Lett. **91**, 107902 (2003).
[18] M. Friesen, P. Rugheimer, D. E. Savage, M. G. Lagally, D. W. van der Weide, R. Joynt and M. A. Eriksson, Phys. Rev. B **67,** 121301(R) (2003).
[19] P. Solinas, P. Zanardi, N. Zanghi and F. Rossi, Phys. Rev. B **67**, 121307(R) (2003).
[20] J. H. Jefferson, M. Fearn and D. L. J. Tipton, Phys. Rev. A **66**, 042328 (2002).
[21] V. N. Golovach and D. Loss, Semicond. Sci. Tech. **17**, 355 (2002).
[22] T. D. Ladd, J. R. Goldman, F. Yamaguchi and Y. Yamamoto, Phys. Rev. Lett. **89**, 017901 (2002).
[23] B. Zeng, D. L. Zhou, Z. Xu, C. P. Sun and L. You, Phys. Rev. A **71**, 022309 (2005).
[24] D. A. Lidar, I. L. Chuang and K. B. Whaley, Phys. Rev. Lett. **81**, 2594 (1998).
[25] D. A. Lidar and K. B. Whaley, quant-ph/0301032 (Jan 2003).
[26] A. Lyakhov and C. Bruder, New J. Phys. **7**, 181 (2005).
[27] T. P. Orlando, J. E. Mooij, L. Tian, C. H. van der Wal, L. Levitov, S. Llyod and J. J. Mazo, Phys. Rev. B **60**, 15398 (1999).
[28] A. Lyakhov and C. Bruder, quant-ph/0609235 (Sep 2006).
[29] A Shnirman, G. Schön and Z. Hermon, Phys. Rev. Lett. **79**, 2371 (1997).
[30] D. V. Averin, Solid State Commun. **105**, 659 (1998).
[31] M. F. Bocko, A. M. Herr and M. J. Feldman, IEEE Tran. Appl. Supercond. **7**, 3638 (1997).
[32] J E. Mooij et al., Science **285**, 1036 (1999).
[33] D. V. Averin, J. R. Friedman and J. E. Lukens, Phys. Rev. B **62**, 11802 (2000).
[34] J. Clarke et al., unpublished. T. V. Filippov et al., unpublished. J. E. Mooij, unpublished
[35] Y. Makhlin, G. Schön and A. Shnirman, Nature. **398**, 305 (1999).
[36] Y. Makhlin, G. Schön and A. Shnirman, Rev. Mod. Phys. **73**, 357 (2001).
[37] P. K. Gagnebin, S. R. Skinner, E. C. Behrman, J. E. Steck, Z. Zhou and S. Han, Phys. Rev. A **72**, 042311 (2005).
[38] W. K. Wootters and W. Zurek, Nature **299**, 802 (1982).
[38] P. Jonker and J. Han, "On Quantum and Classical Computing with Arrays of Superconducting Persistent Current Qubits", IEEE 2000.